\documentclass[twocolumn,aps,prd,amsmath,superscriptaddress]{revtex4-1}
\usepackage{setspace}
\usepackage{color}
\usepackage{fancyhdr}
\usepackage{graphicx}
\usepackage[ansinew]{inputenc}
\usepackage{amssymb}
\usepackage{amsmath}
\usepackage{cancel}
\usepackage{float}
\usepackage{graphicx}
\usepackage{float}
\usepackage{todonotes}
\usepackage{subfigure}
\begin{document}
%===============================================================================
\title{Stable hybrid stars within a SU(3) Quark-Meson-Model}
%===============================================================================
%==============================-Autoren-========================================
%===============================================================================
\author{Andreas Zacchi}
\email{zacchi@astro.uni-frankfurt.de}
% \homepage{www.uni-frankfurt.de/~zacchi}
\affiliation{Institut für Theoretische Physik, Goethe Universität Frankfurt, 
Max von Laue Strasse 1, D-60438 Frankfurt, Germany}

\author{Matthias Hanauske}
\email{hanauske@astro.uni-frankfurt.de}
\affiliation{Institut für Theoretische Physik, Goethe Universität Frankfurt, 
Max von Laue Strasse 1, D-60438 Frankfurt, Germany}
\affiliation{Frankfurt Institute for Advanced Studies, Ruth-Moufang-Strasse 1, 60438 Frankfurt, Germany}

\author{Jürgen Schaffner-Bielich}
\email{schaffner@astro.uni-frankfurt.de}
% \homepage{www.astro.uni-frankfurt.de/~schaffne/}
\affiliation{Institut für Theoretische Physik, Goethe Universität Frankfurt, 
Max von Laue Strasse 1, D-60438 Frankfurt, Germany}
%===============================================================================
\date{\today}
%===============================================================================
   %============================================================================
   \begin{abstract}
   %============================================================================
The inner regions of the most massive compact stellar objects might be occupied by a phase of quarks.
Since the observations of the massive pulsars PSR J1614-2230 and of PSR J0348+0432 with about two solar 
masses, the equations of state constructing relativistic stellar models have to 
be constrained respecting these new limits. We discuss stable hybrid stars, i.e.
compact objects with an outer layer composed of nuclear matter and with a core 
consisting of quark matter (QM).
For the outer nuclear layer we utilize a density dependent nuclear equation of 
state and we use a chiral SU(3) 
Quark-Meson model with a vacuum energy pressure to describe the objects core. 
The appearance of a disconnected mass-radius branch emerging from the hybrid 
star branch
implies the existence of a third familiy of compact stars, so called 
\textit{twin} stars.
Twin stars did not emerge as the transition pressure
has to be relatively small with a large jump in energy density, 
which could not be satisfied within our approach. This is, among other reasons,
due to the fact that the speed of sound in QM has to be relatively high, which 
can be accomplished by an increase of the repuslive coupling. This increase on 
the other hand yields too high transition pressures for twins stars to appear.
% ====================================================
\end{abstract}
\maketitle
%===============================================================================
%-=====================   The sections   ======================================
  %=============================================================================
  \section{Introduction}
  %=============================================================================
A proto-compact star is formed in the aftermath of a supernova explosion, 
which is one of the most extreme events to occur in the universe. At low 
temperature and finite baryon density these objects contain the densest matter 
known to mankind, which exceeds even nuclear density 
($\rho_0 \approx 2.5 \cdot 10^{14}$ g/cm$^3$).The recent measurements of the 
massive pulsars PSR J1614-2230 \cite{Demorest:2010bx} and of PSR J0348+0432 
\cite{Antoniadis:2013pzd} with about two solar masses exceed the highest 
measured pulsar mass of PSR 1913+16 with $M \sim 1.44 M_{\odot}$ by far 
\cite{Hulse75}. This new mass limit sets constraints on the equation of state of
dense matter within compact stellar objects. The repulsive effect of the strong
interaction triples the maximum obtainable mass compared to a non-interacting 
Fermi gas of neutrons \cite{Zdunik:2012dj}. An appropriate EoS therefore should 
yield solutions for compact stars with $\gtrsim 2M_{\odot}$ and illustrates 
likewise the importance of the incorporated interactions.\\
Spherically symmetric compact stars are generally described by 
the Tolman-Oppenheimer-Volkoff equations (TOV equations) \cite{Tolman34}. 
These equations can be derived by solving the Einstein field equations
   \begin{equation}\label{eq:field_eqn}
 G_{\mu\nu}=R_{\mu\nu}-\frac{1}{2}R g_{\mu\nu}=-\frac{8\pi G}{c^4}T_{\mu\nu}
   \end{equation}
where $G_{\mu\nu}$ is the Einstein tensor, $R_{\mu\nu}$ a contraction of the 
Rieman curvature tensor, called Ricci tensor, R being the curvature scalar and 
$T_{\mu\nu}$ the energy-momentum tensor of a relativistic fluid. Under the above
mentioned assumptions one arrives at
   \begin{eqnarray} \label{eq:tov_1}
\frac{dm}{dr}&=&4\pi r^2 \epsilon (r)\\ \label{eq:tov_2} 
\frac{dp}{dr}&=&-\frac{\epsilon(r)m(r)}{r^2}\\ \nonumber
&\cdot&\left(1+\frac{p(r)}{\epsilon(r)}\right)\left(1+
\frac{4\pi r^3p(r)}{m(r)}\right)\left(1-\frac{2 m(r)}{r}\right)^{-1}
   \end{eqnarray}
in units where $c=G=1$.\\   
The solutions of these equations are determined by different equations of state
(EoS), and the entire collection of masses and corresponding radii is called the
mass-radius 
relation (MR) of compact stars \cite{Sagert:2005fw}. 
For each EoS, $p(\epsilon)=p(\epsilon(r))$, where p is the pressure and 
$\epsilon$ the corresponding energy density at a given radius r, exists a 
solution which is parametrized by $p_c$, the central pressure of the star.\\
Two different types of compact stars containing quark matter ought to be 
considered. The first one is based on the idea that the appearence of the 
strange quark lowers the energy per baryon and consequently forms the true 
ground state of nuclear matter, i.e. forms the whole star 
\cite{Itoh:1970uw,Bodmer:1971we,Witten:1984rs}. The resulting object is 
called a pure quark star and has been entirely discussed within the SU(3) model 
in \cite{Zacchi:2015lwa}.
The second one is called a hybrid star, where a quark matter (QM) core is 
surrounded by an outer crust of hadronic matter (HM). The transition from 
nuclear matter to quark matter can occur either in a mixed phase (Gibbs 
construction) or, assuming that there exists a first order 
phase transition at $p_t$, at a sharp transition (Maxwell construction).\\ 
Now, the particle transformations described by the EoS may influence the compressibility of 
the star, which can affect the stability. Is this effect 
significantly enough to alter the properties of the resulting 
compact object, i.e. give rise to a third family of degenerate stars, so called \textit{twin} stars?
These objects would again be stable at a smaller radius but similar mass as the former compact star.
A possible evidence of twin stars goes along with a discontinuity in the EoS.
\cite{Kaempfer:1981a,Kaempfer81,Kaempfer82,Kaempfer83a,Kaempfer83b,Kaempfer85,Glendenning:1998ag,Schertler:2000xq,Blaschke:2015uva}.  
In this article we study various EoS and their solutions within the TOV 
equations using a Maxwell construction. A stable hybrid star configuration with $p_c \geq p_t$ 
is given, if the mass of the star continues to increase after the quark matter 
core appears \cite{Alford:2014dva,Alford:2015dpa, Alford:2015gna}. 
As soon as the mass decreases with larger central pressures, the configurations become unstable.
If the mass then, after decreasing, increases again with 
larger $p_c$, a stable twin star configuration would have been established.
This behaviour is determined by the energy discontinuity $\Delta\epsilon$ 
between the two EoS and the speed of sound within the object.
The works of Alford et. al \cite{Alford:2014dva,Alford:2015dpa, Alford:2015gna} 
confirmed that a stable connected hybrid star branch
emerges from the hadronic branch if the energy density discontinuity is less 
than a critical value. They used a 
constant speed of sound parametrization within the fields correlator 
method for the QM EoS to provide a general framework for empirical testing
and comparison. The recent observations of the $2M_{\odot}$-stars
\cite{Demorest:2010bx,Antoniadis:2013pzd} constraints 
the constant speed of sound parametrization. A stiffer HM EoS and $c_s^2 \geq \frac{1}{3}$ 
for the QM EoS yields solutions with star sequences 
$\geq 2M_{\odot}$ in their approach. In this work we will work with a density dependent (DD2) 
nuclear matter EoS \cite{Typel:2009sy} for the outer layers of the star and a 
chiral SU(3) EoS derived from the Quark-Meson model \cite{Zacchi:2015lwa} for 
the stars core. In \cite{Zacchi:2015lwa} pure quark star 
configurations $\geq 2M_{\odot}$ for a small parameter rage were found, in this model all other solutions
were hybrid stars completely buildt of a mixed phase of HM and QM. 
We scan the same parameters of the SU(3) EoS as in \cite{Zacchi:2015lwa} 
to look for possible twin stars emerging from a stable hybrid star.
  %=============================================================================
  \section{The models}
  %=============================================================================
According to lattice QCD 
calculations, the phase transition at high baryonic densities is of first order 
\cite{Fodor:2001pe,Fodor:2007ue,Alvarez-Castillo:2013cxa}. 
Based on this assumption the transition from hadronic matter to quark matter
is described via a Maxwell construction 
\cite{Glendenning:1992vb,Bhatta04,Bhattacharyya:2009fg}. 
The quark-meson model couples mesons as mediators of the strong interaction 
to quarks utilizing chiral symmetry \cite{Koch:1997ei} via a Yukawa type 
coupling. The coupled equations of motions of the meson fields derived from the 
grand canonical potential have to be solved self-consistently and determine 
finally the EoS. Possible resulting pure quark stars emerging from the chiral 
SU(3) Quark Meson model have been discussed entirely in \cite{Zacchi:2015lwa} 
such as the derivation of the EoS.
  %=============================================================================
  \subsubsection{Chiral Quark Meson Model}
  %=============================================================================
The SU(3) Lagrangian $\mathcal{L}$ of the chiral quark-meson model reads
\begin{eqnarray}
\label{qmlagrangian} 
\mathcal{L}&=&\bar{\Psi}\left(i\cancel{\partial}-
g_\varphi\varphi-g_v\gamma^\mu V_\mu\right)\Psi \\ \nonumber
&+&tr(\partial_{\mu}\varphi)^{\dagger}(\partial^{\mu}\varphi)+
tr(\partial_{\mu} V)^{\dagger}(\partial^{\mu} V)\\ \nonumber
&-&\lambda_1[tr(\varphi^{\dagger}\varphi)]^2-\lambda_2
tr(\varphi^{\dagger}\varphi)^2\\ \nonumber
&-&m_0^2(tr(\varphi^{\dagger}\varphi))-m_v^2 (tr(V^{\dagger}V))\\ \nonumber
&-&tr[\hat{H}(\varphi+\varphi^{\dagger})]+c\left(\det(\varphi^{\dagger})+
\det(\varphi)\right)
\end{eqnarray}
for $SU(3)\times SU(3)$ chiral symmetry incorporating the scalar ($\varphi$) and
vector ($V_\mu$) meson nonet.  Here, $m_v$ stands for the vacuum mass of the 
vector mesons and $\lambda_1$, $\lambda_2$, $m_0$, and $c$ are the standard 
parameters of the linear $\sigma$ model 
\cite{Torn97,Koch:1997ei,Lenaghan:2000ey,Parganlija:2012gv}. 
The matrix $\hat{H}$ describes the explicit breaking of chiral symmetry. The 
quarks couple to the meson fields via Yukawa-type interaction terms with the 
couplings strengths $g_\varphi$ and $g_v$ for scalar and vector mesons, 
respectively.\\ 
The energy density and the pressure are then determined to
    \begin{eqnarray} 
  \epsilon&=&\epsilon_e+\frac{\lambda_1}{4}(\sigma_n^2+\sigma_s^2)^2+
  \frac{\lambda_2}{4}(\sigma_n^4+\sigma_s^4)\\ \nonumber
  &+&\frac{m_0^2}{2}(\sigma_n^2+\sigma_s^2)-\frac{2\sigma_n^2\sigma_s}{\sqrt{2}}
  \cdot c-h_n\sigma_n-h_s\sigma_s+B\\ \nonumber
  &+&\frac{1}{2}\left(m_{\omega}^2\omega^2+m_{\rho}^2\rho^2+
  m_{\phi}^2\phi^2\right)\\ \nonumber
  &+&\frac{3}{\pi^2}\sum_{f=u,d,s}\int_0^{k_F^f}dk\cdot k^2\left(\sqrt{k_{n,s}^2
  +\tilde{m}^2}\right)
    \end{eqnarray}
and
    \begin{eqnarray} \label{potV} 
  p&=&-\frac{1}{2}\left(m_{\omega}^2\omega^2+m_{\rho}^2\rho^2+m_{\phi}^2\phi^2
  \right)\\ \nonumber 
  &+&\frac{\lambda_1}{4}(\sigma_n^2+\sigma_s^2)^2+\frac{\lambda_2}{4}(\sigma_n^4
  +\sigma_s^4)\\ \nonumber 
  &+&\frac{m_0^2}{2}(\sigma_n^2+\sigma_s^2)-\frac{2\sigma_n^2\sigma_s}{\sqrt{2}}
  \cdot c-h_n\sigma_n-h_s\sigma_s+B \\ \nonumber
  &+&\frac{3}{\pi^2}\sum_{f=u,d,s}\int_0^{k_F^f}dk\cdot k^2\left(\sqrt{k_{n,s}^2
  +\tilde{m}^2}-\tilde{\mu}_f\right)
    \end{eqnarray}
where the indices n$=$nonstrange (up- and down quarks) and s$=$strange quarks. 
For the coulings and masses of
the included fields standard values are assumed. A detailled treatment on the 
parameters can be found in 
\cite{Koch:1997ei,Lenaghan:2000ey,Schaefer:2008hk,Zacchi:2015lwa}.
Since the properties of the reviewed hybrid stars depends only on the parameters
of the quark sector, a broader overview shall be given compared to the nuclear 
matter parameter range. However, four parameters can be varied: 
   \begin{enumerate}
 \item The constituent quark mass $m_q$ determines the scalar coupling  for the 
       nonstrange $g_n$ and strange condensate $g_s$ via the Goldberger-Treiman 
       relation: $g_n=\frac{m_q}{f_{\pi}}$ and $g_s=g_n\sqrt{2}$, where $g_s$ is
       adoped from SU(3) symmetry considerations.
 \item The vector coupling is independent of the constituent quark mass, it will 
       be varied in the scale of the scalar coupling, $g_{\omega}\sim g_n$,  to 
       study its influences in an appropriate range. The strange coupling of the
       $\phi$-meson is fixed by SU(3) symmetry.
 \item The experimentally not well determined mass of the $\sigma$-meson covers 
       a range from $400$~MeV $\leq m_{\sigma} \leq 800$~ MeV 
       \cite{Parganlija:2010fz,Agashe:2014kda}.
 \item The Bag constant B models the confinement and can be 
       interpreted as a vacuum energy density term. The fields are independent 
       of its variation, its impact is to stiffen or soften the EoS. Physically 
       reasonable ranges within this context are 
       $60$~MeV $\leq B^{\frac{1}{4}} \leq 200$~MeV.
   \end{enumerate}
  %=============================================================================
  \section{Hybrid Stars}
  %=============================================================================
% von Schertler/Leupold
At large densities hadronic matter is expected to undergo two phase 
transitions. The first one deconfines hadrons to quarks and gluons. 
Note that in a strict sense neither the deconfinement phase transition nor the 
chiral phase transition can be described by an order parameter based on
underlying symmetries of QCD.
The second one restores chiral symmetry. Yet it is
an unsettled issue wheter these transitions are real phase transitions or 
crossover transitions \cite{Schertler:1999xn}. 
We will study and compare various models at ultrahigh densities 
to search for differences and similarities as well as their resulting 
predictions for compact objects, i.e. the mass-radius relation.
  %=============================================================================
  \subsection{Construction of the phase transition}
  %============================ ================================================
The study of the deconfined phase transition is related to the mixed phase. 
It has been suggested, that the mixed phase in compact objects behaves more in accordance
with the Maxwell construction than with the Gibbs construction 
\cite{Maruyama:2006jk,Bhattacharyya:2009fg,Hempel:2013tfa}.
Furthermore it is more likely that twin stars appear within the Maxwell construction, according to \cite{Bhattacharyya:2009fg}.
In this article we thus utilize a Maxwell construction due to the above mentioned reasons.
In refs \cite{Alford:2014dva,Alford:2015dpa,Alford:2015gna} the QM EoS was parametrized
a relatively simple 
form (see eq.~\ref{qmeostwo}) and the transition from HM to QM can be constructed
without any constraints concerning the chemical potential. Our approach on the 
other hand needs to take into account the pressure as function of the chemical 
potential to find the thermodynamically justified 
transition pressure (see fig.~\ref{fig:pandmu} and the discussion there).  
\\
In electrically neutral stellar matter baryon number and charge have to be 
conserved quantities. Under this assumption the chemical potential of species i 
can be defined as
    \begin{equation}\label{chem_pot}
  \mu_i = B_i \mu_B + Q_i \mu_Q
    \end{equation}
where $B_i$ is the baryon number and $Q_i$ the charge in units of the electron 
charge and $\mu_B$ and $\mu_Q$ are the baryonic and electric chemical potentials
respectively. Note, that strangeness is not a conserved quantity. The phase 
transition from HM to QM produces a mixed phase. Now, the Gibbs condition 
requires that the coexisting phases have opposite charge and it might also 
happen that the mixed phase is energetically too expensive 
\cite{Bhattacharyya:2009fg,Maruyama:2006jk}. Then the two phases are in direct 
contact with each other, which corresponds to a Maxwell construction, where
    \begin{eqnarray}
  P_{HM}(\mu_B,\mu_Q)&=&P_{QM}(\mu_B,\mu_Q)\\
  \mu_B&=&\mu_{HM}=\mu_{QM}
    \end{eqnarray}
The baryon chemical potential is continuous, but $\mu_Q$ jumps at the interface 
of the two phases, so that the phase transition takes place if the pressure of 
the QM phase equals the pressure of the HM phase at a given baryo-chemical 
potential $\mu_B$. The MC corresponds to 
constant pressure in the energy density interval of the mixed phase, whereas the
pressure increases with baryon density in the GC.\\
However, the existence of a quark phase in a compact star requires the 
transition pressure to be smaller than the central pressure of the star, which 
is valid for the MC and also for the GC. 
  %=============================================================================
  \subsection{Stability Criteria}
  %=============================================================================
As long as the mass of the star is an increasing function of $p_c$ the compact 
object will be stable. Since a hybrid star contains a QM-core, there exists a 
threshold value in the jump in energy density $\Delta\epsilon_c$ which determines
the stars stability when the QM-core first appears. 
    \begin{equation}\label{eq:constraint}
  \frac{\Delta\epsilon_c}{\epsilon_t}=\frac{1}{2}+\frac{3}{2}\frac{p_t}
  {\epsilon_t}
    \end{equation}
where $\epsilon_t$ and $p_t$ are the values of the energy density and pressure 
at the phase transition. For a derivation and discussion of 
(\ref{eq:constraint}) see \cite{seidov:1971pty,Kaempfer:1981a,Kaempfer81,
Kaempfer82,Kaempfer83,Kaempfer83a,zdunikhaenselschaeffer:1983pti,
Lindblom:1998dp}.\\ 
For a high value of $\Delta\epsilon$ the cusp in the MR relation is hardly 
detectable and in the range of $\sim 10^{-4}M_{\odot}$ in agreement with 
\cite{Lindblom:1998dp,Alford:2014dva,Blaschke:2015uva}, i.e. shortly after the 
QM core appears the QM core is unable to counteract the gravitational attraction
from the HM and the star becomes unstable. 
  %=============================================================================
  \section{Results}
  %=============================================================================
The appearence of a QM core within a compact star is entirely determined by the 
transition pressure $p_t$ and the discontinuity in the energy density 
$\Delta \epsilon$. If the pressure within the star lies below the transition 
pressure, the object would be entirely determined by the HM EoS and could not be
classified as a hybrid star. The relation $\frac{\Delta \epsilon}{\epsilon_t}$ 
as a function of $\frac{p_t}{\epsilon_t}$ will become important in context with 
eq.~(\ref{eq:constraint}) when investigating for connected or disconnected 
hybrid star branches \cite{Alford:2014dva, Alford:2015dpa, Alford:2015gna}. 
%===============================================================================
  \subsection{Various EoS and the corresponding mass-radius 
  relations for fixed B and different $g_{\omega}$}
\begin{figure}[H]
\center
\includegraphics[width=\columnwidth]{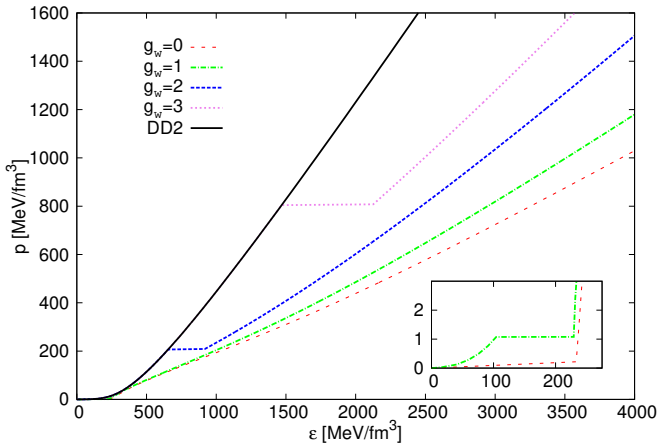}
\caption{\textit{The EoS with fixed $B=60$~MeV while varying $g_{\omega}$ at
    $m_{\sigma}=600$~MeV and $m_q=300$~MeV. The inlaid figure accentuates
    the behaviour of the EoS for $g_{\omega}=0$ and $g_{\omega}=1$ which else is
    hardly perceivable.}}
\label{fig:eos60}
\end{figure}

\begin{figure}[H]
\center
\includegraphics[width=\columnwidth]
{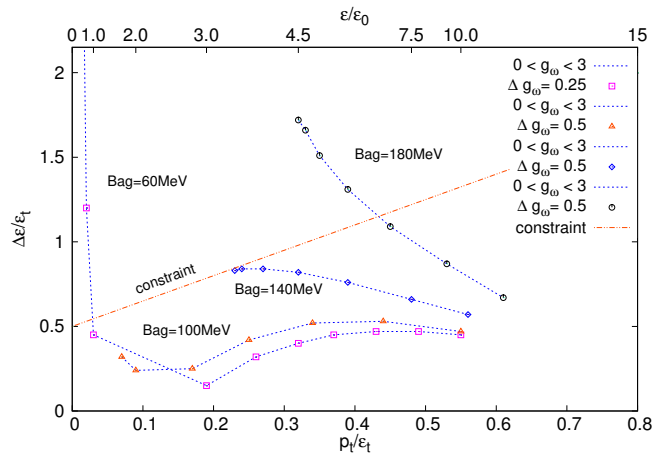}
\caption{\textit{Phase diagram for hybrid stars with fixed 
         B while varying $0 \leq g_{\omega} \leq 3$ at constant
         $m_{\sigma}=600$~MeV and $m_q=300$~MeV. The axes display the transition 
         pressure $p_t$ and the energy density discontinuity $\Delta \epsilon$
         in units of the nuclear energy density at the transition $\epsilon_t$. Note,
         that the first data point for the $B=60$~MeV line (on the left) corresponds to
         $g_{\omega}=1$. The following data points are incremented by $\Delta g_{\omega}=0.25$.  
         }}
\label{fig:constraintbag}
\end{figure}

  %=============================================================================
Figure~\ref{fig:eos60} shows the total hybrid EoS for a fixed value of the 
vacuum pressure B~$=60$~MeV while varying the vector coupling constant from 
$0 \leq g_{\omega} \leq 3$.\\  
%====== = = = = = = = = = == == = = = = = = = = = = = = = = = ==================
%====== = = = = = = = = = == =      = = = = = = = = = = = = = ==================
For increasing values of the repulsive coupling the transition pressure $p_t$ 
increases due to a stiffening in the QM EoS.\\
The $g_{\omega}=0$ case corresponds to a transition from HM to QM at 
$\frac{\epsilon}{\epsilon_0} \leq 1$. 
% Hier ist e_t=0.1877 und e_t2=233.23, also D_epsilon 233:0.18=1240
A transition occuring below saturation energy density is clearly unphysical and shall 
therefore not be discussed any further 
(see upper x axis in fig~\ref{fig:constraintbag}).\\
For $g_{\omega}=1$ the transition occurs at $p_t\simeq 1.5$~$MeV/fm^3$ and 
$\epsilon_t \simeq 102$~$MeV/fm^3$ (see inlaid figure in fig.~\ref{fig:eos60}). 
The discontinuity in energy density here
is $\Delta \epsilon \simeq 122$~$MeV/fm^3$. In this case 
$\frac{\epsilon}{\epsilon_0} \simeq 1$, see also fig.~\ref{fig:constraintbag}, 
which corresponds to the leftmost data point on the $g_{\omega}=0$ line. 
Note that in fig.~\ref{fig:eos60} and in all following graphics the pure 
HM results are shown as a reference, denoted as ``DD2''. 
%====== = = = = = = = = = == == = = = = = = = = = = = = = = = ==================
%====== = = = = = = = = = == =  = = = = = = = = = = = = = = = ==================
\begin{figure}[H]
\center
\includegraphics[width=\columnwidth]
{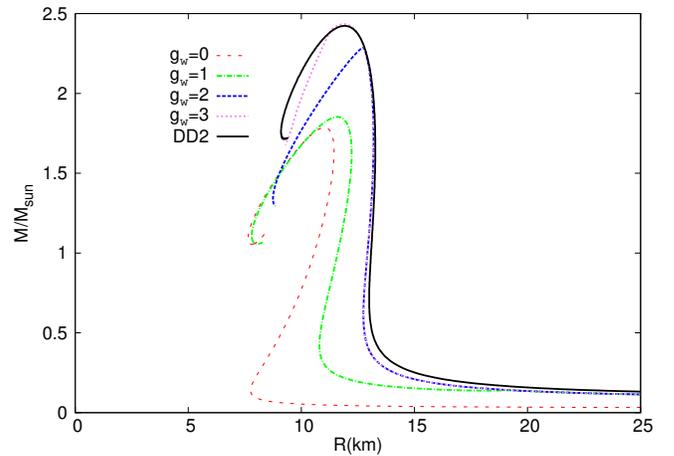}
\caption{\textit{The mass-radius relation with fixed $B=60$~MeV while varying 
                 $g_{\omega}$ at $m_{\sigma}=600$~MeV and $m_q=300$~MeV.}}
\label{fig:mr60}
\end{figure}
%====== = = = = = = = = = == = = = == = = = = = = = = = = = = ==================
The corresponding mass-radius relation is shown in fig. \ref{fig:mr60}. 
For $g_{\omega}=1$ the phase transition from HM to QM does not destabilize the 
star for a relatively wide range in mass, i.e. the emerging QM core gets larger while 
the hybrid star manages to stay stable up to $\sim 1.7 M_{\odot}$. This 
behaviour is very similar to the one of the hadronic mode ``DD2'', but shifted 
to smaller masses and radii.\\
A repulsive coupling of $g_{\omega}=2$ on the other hand results in 
a connected hybrid star branch hardly detectable compared to
$g_{\omega}=1$ and with a similar trend as the ``DD2'' case, but with solutions 
reaching $\gtrsim 2 M_{\odot}$.\\
For $g_{\omega}=3$ the transition sets in at already unstable configurations 
for the pure nuclear matter case.
%====== = = = = = = = = = == = = == = = = = = = = = = = = = = ==================
\begin{figure}[H]
\center
\includegraphics[width=\columnwidth]
{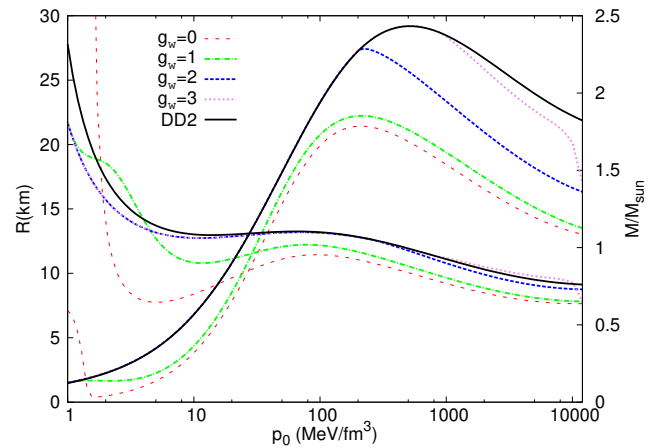}
\caption{\textit{The radius- and mass curves as fuction of $p_c$ with fixed 
                 $B=60$~MeV while varying $g_{\omega}$  at $m_{\sigma}=600$~MeV 
                 and $m_q=300$~MeV. The curves starting in the upper left region
                 are the radius curves whereas the curves starting on the lower 
                 left side are the mass curves.}}
\label{fig:cp60}
\end{figure}
%====== = = = = = = = = = == = == = = = = = = = = = = = = = = ==================
Figure~\ref{fig:cp60} displays the radius- and mass curves as fuction of 
$p_c$ with $B\geq 60$~MeV while varying $g_{\omega}$  at $m_{\sigma}=600$~MeV and 
$m_q=300$~MeV. The curves starting in the upper left region are the radius 
curves for  a given value of $g_{\omega}$. The curves starting on the lower left 
side are the mass curves. The associated x-axis in fig.~\ref{fig:cp60} shows the
pressure pertaining to both curves. 
The curves leave the hadronic ``DD2'' reference line at the respective transition pressure $p_t$ 
and, still rising, yielding stable hybrid star solutions. Unstable solutions can be read off 
from the point on where the mass decreases with increasing pressure.
These features are valid for all following radius- and mass curves as 
fuction of $p_c$.\\
Figure~\ref{fig:cp60} substantiates the hithero discussion regarding the 
increase of the repulsive coupling by depicting up to which central pressure 
the hybrid star configurations stay stable: 
With higher repulsive coupling, the appearing hybrid star configurations become 
unstable, i.e. the smaller the resulting QM core, though the masses are significantly higher.

%====== = = = = = = = = = = = = = = = = = = = = = = = = = = = ==================
%====== = = = = = = = = = = = = = = = = = = = = = = = = = = = ==================
\begin{figure}[H]
\center
\includegraphics[width=\columnwidth]
{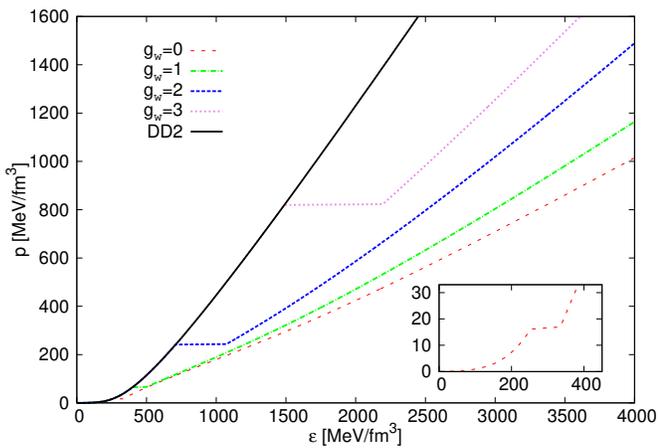}
\caption{\textit{The EoS with fixed $B=100$~MeV while varying $g_{\omega}$ at
    $m_{\sigma}=600$~MeV and $m_q=300$~MeV. The inlaid figure shows
    the behaviour of the EoS for $g_{\omega}=0$. }}
\label{fig:eos100}
\end{figure}
\begin{figure}[H]
\center
\includegraphics[width=\columnwidth]
{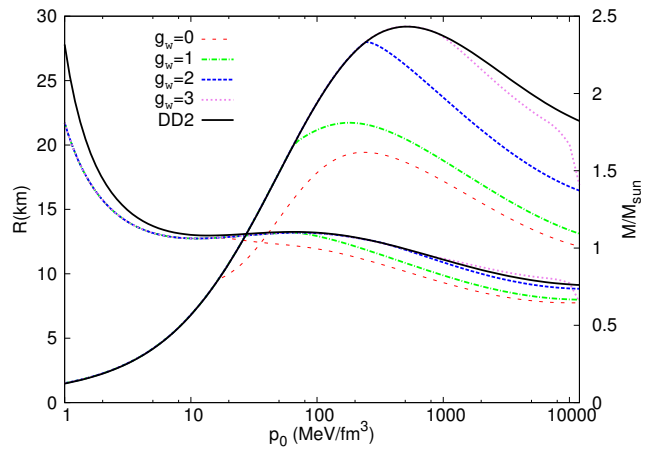}
\caption{\textit{The radius- and mass curves as fuction of $p_c$ with fixed 
                 $B=100$~MeV while varying $g_{\omega}$  at $m_{\sigma}=600$~MeV
                 and $m_q=300$~MeV. The curves starting in the upper left region
                 are the radius curves whereas the curves starting on the lower 
                 left side are the mass curves.}}
\label{fig:cp100}
\end{figure}

%====== = = = = = = = = = == = =  = = = = = = = = = = = = = = ==================
Figure~\ref{fig:eos100} shows the EoS for $B=100$~MeV.
The transition pressure increases with an associate increase of 
the jump in energy density.\\ 
For $g_{\omega}=0$ with fixed $B=100$~MeV the respective values are 
$p_t \simeq 15$~$MeV/fm^3$, $\epsilon_t \simeq 230$~MeV and 
$\Delta \epsilon \simeq 90$~$MeV/fm^3$, see inlaid figure in figure \ref{fig:eos100} 
and see fig.~\ref{fig:constraintbag} for 
$\frac{\epsilon}{\epsilon_0} \simeq 1.8$ respectively.\\
For $g_{\omega}=1$ and $B=100$~MeV we find $p_t \simeq 75$~$MeV/fm^3$, 
$\epsilon_t \simeq 380$~$MeV/fm^3$ and $\Delta \epsilon \simeq 100$~$MeV/fm^3$ at 
$\frac{\epsilon}{\epsilon_0} \simeq 2.8$, see also fig.~\ref{fig:constraintbag}.
%====== = = = = = = = = = == = = = =  = = = = = = = = = = = = ==================
%====== = = = = = = = = = = = = = = = = = = = = = = = = = = = ==================
\begin{figure}[H]
\center
\includegraphics[width=\columnwidth]
{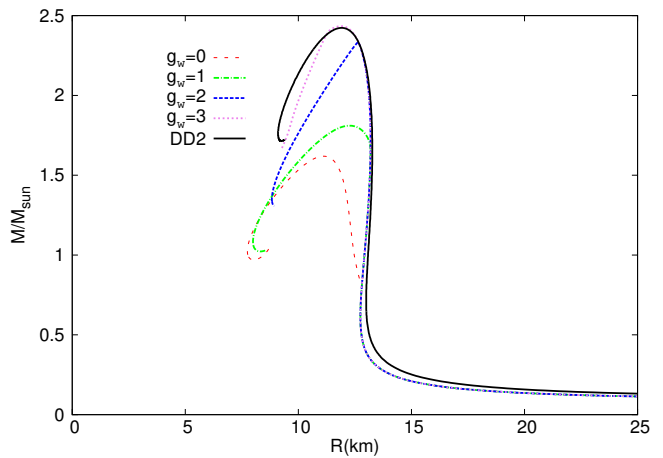}
\caption{\textit{The mass-radius relation with fixed $B=100$~MeV while varying 
                 $g_{\omega}$ at $m_{\sigma}=600$~MeV and $m_q=300$~MeV.}}
\label{fig:mr100}
\end{figure}
%====== = = = = = = = = = == = = = =  = = = = = = = = = = = = ==================
The resulting mass-radius relations for these EoS are shown in 
fig.~\ref{fig:mr100}. Increasing further the repulsive coupling leads to hybrid 
star configurations, which do not support a stable QM core ($g_{\omega} \geq 2$). 
The trends of the curves obviously show differences to the $B=60$~MeV parameter choice.
The transiton pressures for $B=100$~MeV are higher compared to the $B=60$~MeV case,
see figs.~\ref{fig:cp60} and \ref{fig:cp100}, and the appearing QM core does
destabilize the configurations. 

%====== = = = = = = = = = == = = == = = = = = = = = = = = = = ==================
\begin{figure}[H]
\center
\includegraphics[width=\columnwidth]{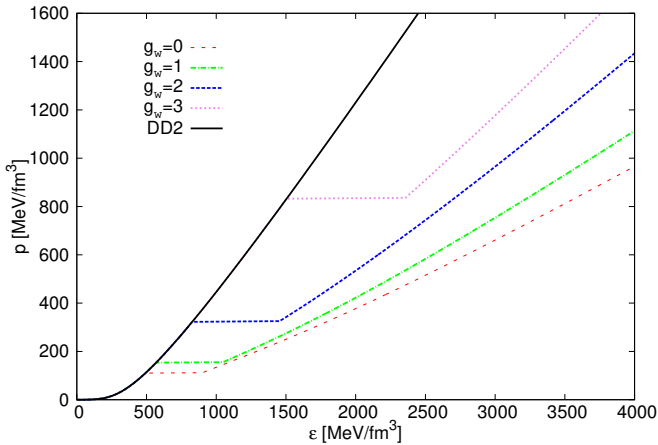}
\caption{\textit{The EoS with fixed $B=140$~MeV while varying $g_{\omega}$ at
    $m_{\sigma}=600$~MeV and $m_q=300$~MeV.}}
\label{fig:eos140}
\end{figure}

%====== = = = = = = = = = == = = == = = = = = = = = = = = = = ==================
The QM core for $g_{\omega}=0$ appears at $\sim 0.8 M_{\odot}$ at a radius of 
$\sim 12.5$~km, see fig.~\ref{fig:cp100} where the mass and radius lines leaves the 
hadronic ``DD2'' reference line. The star does not get unstable up to $\sim 1.6 M_{\odot}$ at a 
radius of $\sim 11$~km.\\
The QM core for $g_{\omega}=1$ appears at $\sim 1.6 M_{\odot}$. The hybrid
star configurations stay stable up to $\sim 1.7 M_{\odot}$, see 
figs.~\ref{fig:mr100} and \ref{fig:cp100}. The appearance 
of the QM core at $g_{\omega}=1$ destabilizes the star configurations faster 
than in the $g_{\omega}=0$ case for $B=100$~MeV. 
The EoS for $B=140$~MeV is shown in fig.~\ref{fig:eos140}. It shows an 
increase of the transition pressure $p_t$ as expected.
%====== = = = = =  = == = = = = = = = = = = = = = = = = = = = ==================
%====== = = = = = = = = = == = = = = == = = = = = = = = = = = ==================
The discontinuity in energy density increases too, but 
displays a nontrivial relation to $p_t$ which can be observed in greater 
detail in the phase diagram shown in fig.~\ref{fig:constraintbag}.
%====== = = = = = = = = = == = = = = = = = = = = = = = = = = ==================
\begin{figure}[H]
\center
\includegraphics[width=\columnwidth]
{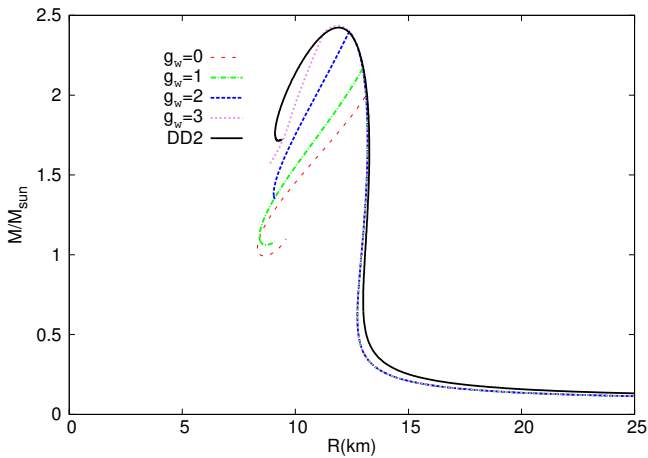}
\caption{\textit{The mass-radius relation with fixed $B=140$~MeV while varying 
                 $g_{\omega}$ at $m_{\sigma}=600$~MeV and $m_q=300$~MeV.}}
\label{fig:mr140}
\end{figure}
%====== = = = = = =         = = = = = = = = = = = = = = = = = ==================
The resulting mass-radius curve for $B=140$~MeV is shown in
fig.~\ref{fig:mr140}. A hybrid star branch appears but is hardly noticable.
As already mentioned, the transitions for a value of $g_{\omega}=3$ sets in at 
an already unstable configurations, i.e. no stable hybrid star branch
at all emerges.
%====== = = = = = = = = = = = = = = = = = = = = = = = = = = = ==================
\begin{figure}[H]
\center
\includegraphics[width=\columnwidth]
{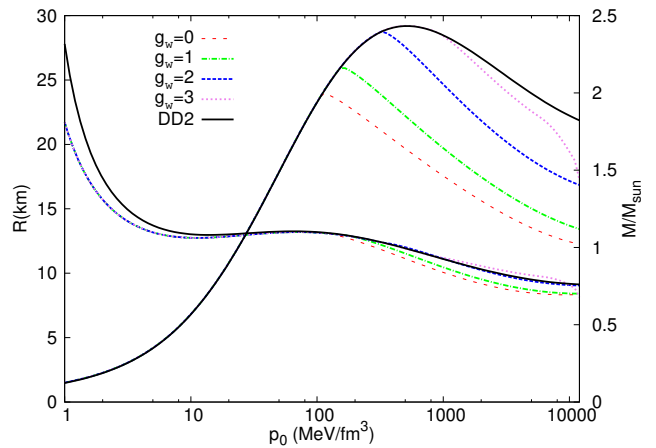}
\caption{\textit{The radius- and mass curves as fuction of $p_c$ with fixed 
                 $B=140$~MeV while varying $g_{\omega}$  at $m_{\sigma}=600$~MeV
                 and $m_q=300$~MeV. The curves starting in the upper left region
                 are the radius curves whereas the curves starting on the lower 
                 left side are the mass curves.}}
\label{fig:cp140}
\end{figure}
%====== = = = = = = = = = = = = = = = = = = = = = = = = = = = ==================
Fig.~\ref{fig:cp140} shows the corresponding radius- and mass curve
as function of the central pressure. The hybrid star configurations follow the 
``DD2'' curve, and become unstable nearly immediately after the appearance of 
the QM core. The repulsive force in the QM EoS is not strong enough to support a
large hadronic mantle. The star would collapse having a too large QM core.\\ 
Generally speaking: Raising the value of the vacuum pressure leads to hybrid 
star branches where the hybrid stars destabilize faster after 
the appearence of the QM core, and the transition occurs at higher masses.
%===============================================================================
%====== = = = = = = = = = == = = = == = = = = = = = = = = = = ==================
The phase diagram displayed in figure~\ref{fig:constraintbag} depicts the ratio 
of pressure to energy density at the transition of hadronic matter
versus the discontinuity in energy density at the transition. 
The upper x axis displays the corresponding central energy density in units of 
nuclear energy density $\epsilon_0 \simeq 145$~$\frac{MeV}{fm^3}$.\\
The transition for small values of B and $g_{\omega}$ occurs at a too small 
central energy density $\frac{\epsilon_t}{\epsilon_0} \leq 1$. For large values of
B and a small repulsive coupling the transition occurs at $4-10$ times nuclear 
saturation density. Within the range $100 \leq B \leq 140$~MeV the transition 
for zero repulsion stays below the constraint line, given by 
eqn.~\ref{eq:constraint}. It is interesting to note that all curves converse 
in an area at around $0.55 \leq \frac{p_t}{\epsilon_t} \leq 0.65$ and 
$0.4 \leq \frac{\Delta \epsilon}{\epsilon_t} \leq 0.6$ where the central energy 
density is $\sim 10$ times nuclear saturation density (even for higher values of
$g_{\omega}$ not displayed here). 
{Figure~\ref{fig:pandmu} displays the pressure as a function of the chemical 
potential $\mu$ for the parameter choice $m_{\sigma}=600$~MeV, $m_q=300$~MeV 
and $B=100$~MeV while varying $0 \leq g_{\omega} \leq 3$.
The intersecting point between the 
HM-and the QM curve indicates where the transition pressure for a given choice
of parameters is located. 
%====== = = = = = = = = = == = = = = = = = = = = = = = = = = = == = = = = = = = 
\begin{figure}[H]
\center
\includegraphics[width=\columnwidth]{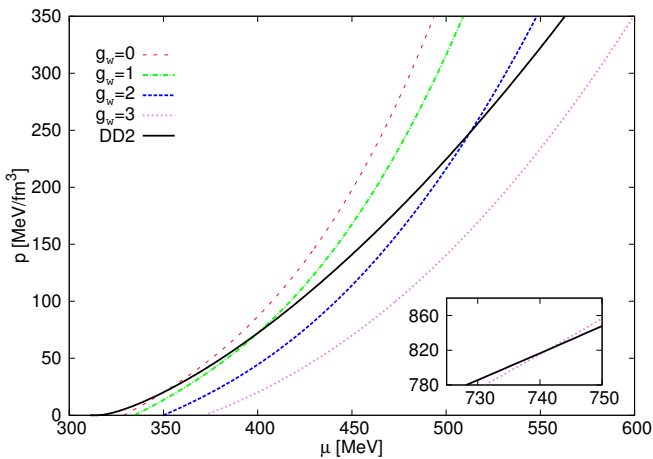}
\caption{\textit{The intersection in the pressure versus chemical potential 
                 $\mu$ plane for $0 \leq g_{\omega} \leq 3$ within the parameter
                 choice $m_{\sigma}=600$~MeV, $m_q=300$~MeV and $B=100$~MeV, 
                 corresponding to figs.~\ref{fig:eos100}, \ref{fig:mr100} and 
                 \ref{fig:cp100}. The MC requires that from the intersecting 
                 point on the dominance in the EoS flips, which creates a QM 
                 core within the star at the corresponding pressure.\\
                 The inlaid figure shows the intersection for the $g_{\omega}=3$
                 case, which is out of the plot range.
    }}
\label{fig:pandmu}
\end{figure}
%====== = = = = = = = = = == = = = = = = = = = = = = = = = = = == = = = = = = =
The intersection for $g_{\omega}=0$ takes place at $p \simeq 15$~$MeV/fm^3$ and 
$\mu \simeq 355$~MeV and for $g_{\omega}=1$ at $p \simeq 75$~$MeV/fm^3$ and 
$\mu \simeq 400$~MeV, see also figs.~\ref{fig:eos100} and \ref{fig:cp100}. 
It is interesting to note that the a stiffer EoS has a ``softer'' behaviour in the 
$p-\mu$~plane. Softer means here that for larger values of $g_{\omega}$ both, 
pressure and $\mu$ increase, i.e. the intersection takes place at a higher 
pressure. That corresponds to a transition from HM to QM at a higher 
central energy density in terms of nuclear energy density, see 
figs.~\ref{fig:constraintbag} and \ref{fig:constraintgw} for comparison 
(upper x-axis). 
An appearing QM core destabilizes the star quite soon, and twin star solutions 
are ruled out, since these require a relatively low transition pressure
\cite{Alvarez-Castillo:2013cxa,Blaschke:2015uva}.
%====== = = = = = = = = = == = = = = = = = = = = = = = = = = = == = = = = = = = 
\begin{figure}[H]
\center
\includegraphics[width=\columnwidth]{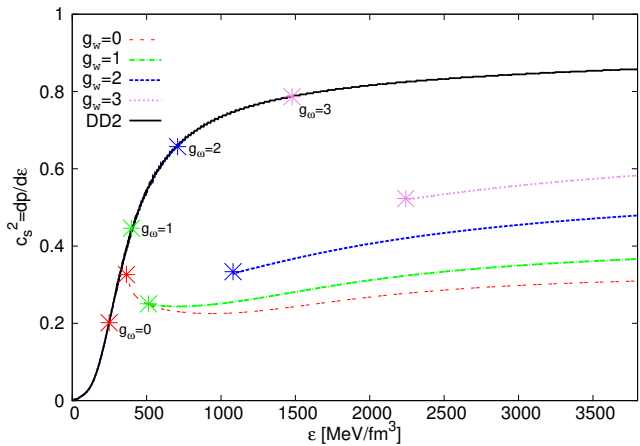}
\caption{\textit{The speed of sound $c_s^2=\frac{dp}{d\epsilon}$ as a function 
                 of the energy density $\epsilon$  for 
                 $0 \leq g_{\omega} \leq 3$ within the parameter choice 
                 $m_{\sigma}=600$~MeV, $m_q=300$~MeV and $B=100$~MeV, 
                 corresponding to figs.~\ref{fig:eos100}, \ref{fig:mr100} and 
                 \ref{fig:cp100}. For this parameter choice the transition is 
                 marked by the symbols on the ``DD2'' curve. 
                 }}
\label{fig:sos}
\end{figure}
% ====================================================
In fig.~\ref{fig:sos} we examine the speed of sound for 
$0 \leq g_{\omega} \leq 3$ within the parameter choice $m_{\sigma}=600$~MeV, 
$m_q=300$~MeV and $B=100$~MeV, corresponding to figs.~\ref{fig:eos100}, 
\ref{fig:mr100} and \ref{fig:cp100}. Since the Bag constant does not affect the 
stiffness of the EoS (it just changes the value of the vacuuum pressure) the 
slope of theses curves for any choice of B would remain the same. Only the 
transition values of the energy density $\epsilon_t$ from one EoS 
to the other EoS would change and in equal steps of $\Delta \epsilon$.\\
For $g_{\omega}=0$ and $g_{\omega}=1$, $\Delta \epsilon \simeq 95$~$MeV/fm^3$, see
also the discussion in the previous sections. The symbols on the ``DD2'' curve 
mark the point where the transition takes place and the stars leave the hadronic 
branch. The corresponding symbols on the QM 
lines mark then the points, where the QM core appears. As one would expect, an increase of the repulsive 
coupling not only stiffens the EoS but also raises the speed of sound within the
medium. The $g_{\omega}=0$ line saturates at $c_s^2=\frac{1}{3}$ which is 
reasonable since ultrarelativistic matter without interactions saturates at 
$p(\epsilon)=\frac{1}{3}\epsilon$ \cite{Shapiro_book,Glendenning_book}.
Since $g_{\omega}=3$ has far too high transition pressures for hybrid- and twin 
stars the highest considered repulsive coupling $g_{\omega}=2$ reaches 
$c_s^2 \simeq 0.5$. That means that all physically relevant and 
considered cases in this work lie within $0.3 \leq c_s^2 \leq 0.5$. This will 
become important in the following when we compare our results with those 
from Alford et. al \cite{Alford:2014dva,Alford:2015dpa, Alford:2015gna}.
   %==============================================================================
   \subsection{Various EoS and the corresponding mass-radius 
   relations for fixed $g_{\omega}$ and different B }%   
   %============================================================================
Figure \ref{fig:eos0} shows the EoS at fixed $g_{\omega}=0$ for various values 
of the Bag constant B. For increasing values of B the transition pressure $p_t$ 
increases. As in the case of increasing B at fixed $g_{\omega}$, increasing B 
while varying $g_{\omega}$ leads to the same behaviour of the different EoS.
%===============================================================================
\begin{figure}[H]
\center
\includegraphics[width=\columnwidth]{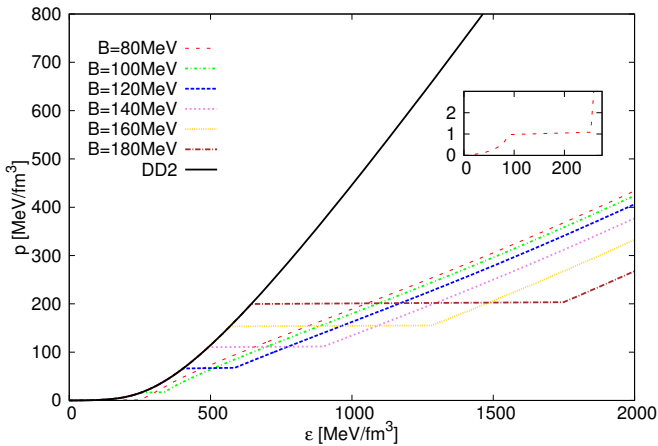}
\caption{\textit{The EoS with fixed $g_{\omega}=0$ while varying B at
                 $m_{\sigma}=600$~MeV and $m_q=300$~MeV.
                 The inlaid figure shows the behaviour of 
                 the EoS for $B=180$~MeV.}}
\label{fig:eos0}
\end{figure}
%====== = = = = = = = = = == = = = =  = = = = = = = = = = = = ==================
For $B=80$~MeV $p_t\simeq 1$~$MeV/fm^3$, $\epsilon_t \simeq 92$~$MeV/fm^3$ and the 
discontinuity in energy density is $\Delta \epsilon \simeq 160$~$MeV/fm^3$ (see inlaid 
figure). For the highest chosen value of $B=180$~MeV $p_t \simeq 202$~$MeV/fm^3$, 
$\epsilon_t \simeq 650$~$MeV/fm^3$ and $\Delta \epsilon \simeq 1100$~$MeV/fm^3$, i.e. the 
discontinuity in the energy density $\Delta \epsilon$ increases also with B.
%====== = = = = = = = = = == = = = = = = = = = = = = = = = = ==================
\begin{figure}[H]
\center
\includegraphics[width=\columnwidth]
{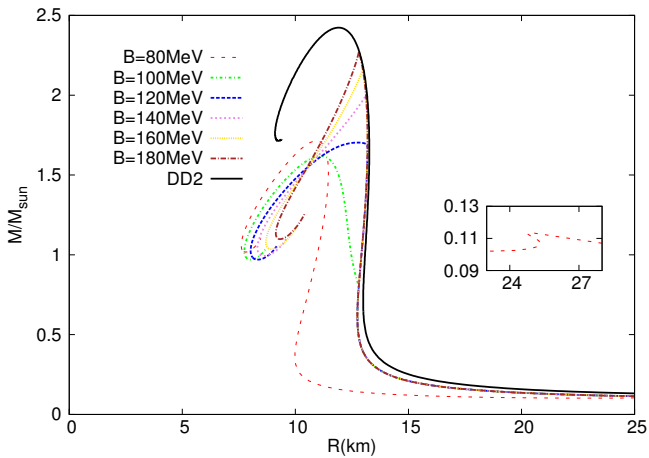}
\caption{\textit{The mass-radius relation with fixed $g_{\omega}=0$ while 
                 varying B at $m_{\sigma}=600$~MeV and $m_q=300$~MeV.
                 The inlaid figure accentuates
    the behaviour of the mass-radius curve for $B=80$~MeV which else is
    hardly perceivable.}}
\label{fig:mr0}
\end{figure}
%====== = = = = = = = = = == = = = =  = = = = = = = = = = = = ==================
Figure~\ref{fig:mr0} shows the mass-radius relations for $g_{\omega}=0$ while 
varying B with $m_{\sigma}=600$~MeV and $m_q=300$~MeV. For the smallest value of 
$B=80$~MeV the QM core appears at already $0.11 M_{\odot}$ at a radius of 
$\sim 25$~km (see inlaid figure), see also fig.~\ref{fig:cp0}. 
The shape of the curve is similar to 
the pure hadronic one but shifted to slightly smaller values of mass and radius due to 
the appearence of the QM core. The transition from HM to QM appears at 
$\frac{\epsilon}{\epsilon_0}\leq 1$, see fig.~\ref{fig:constraintgw}.
The inlaid figure displays a disconnected mass-radius branch, which is 
an indication for a twin star. These disconnected solutions were found up to values 
of $B\simeq 90$~MeV, getting harder to detect with larger B and always at 
physically too small transition energy densities $0.66\leq \frac{\epsilon}{\epsilon_0}\leq 1$, 
see figs.~\ref{fig:constraintgw} and \ref{fig:twin_region}, and shall 
therefore not be discussed any further.\\ 
For $B=100$~MeV the transition occurs at 
$\frac{\epsilon}{\epsilon_0}\simeq 1.8$. The respective values are 
$p_t \simeq 15$~$MeV/fm^3$, $\epsilon_t \simeq 230$~$MeV/fm^3$ and 
$\Delta \epsilon \simeq 90$~$MeV/fm^3$ (see also inlaid figure in 
fig.~\ref{fig:eos100}, fig.~\ref{fig:eos0} and fig.~\ref{fig:cp0}). 
The QM core appears at $\sim 0.8 M_{\odot}$ at a radius of $\sim 12.5$~km. 
The star configuration does 
not get unstable up to $\sim 1.6 M_{\odot}$ at a radius of $\sim 11$~km, which
can altogether be observed in fig.~\ref{fig:cp0}. The resulting mass-radius 
relation for this EoS is also shown in fig.~\ref{fig:mr100}. Higher 
values of the vacuum energy term B lead to much smaller hybrid star branches, 
hardly visible and in accordance with \cite{Alford:2014dva,Alford:2015dpa, Alford:2015gna}. 
The configurations get unstable nearly immediately after the appearence of the 
QM core, which itself emerges at a higher mass.
%====== = = = = = = = = = == =  = = = = = = = = = = = = = = = ==================
\begin{figure}[H]
\center
\includegraphics[width=\columnwidth]
{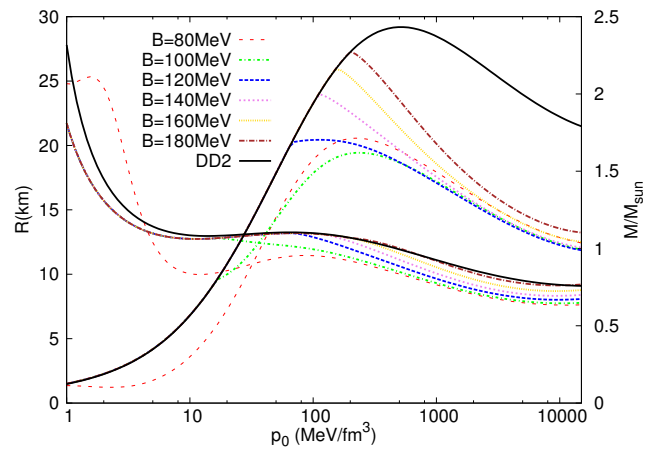}
\caption{\textit{The radius- and mass curves as fuction of $p_c$ with fixed 
                 $g_{\omega}=0$ while varying B at $m_{\sigma}=600$~MeV and 
                 $m_q=300$~MeV. The curves starting in the upper left region
                 are the radius curves whereas the curves starting on the lower 
                 left side are the mass curves.}}
\label{fig:cp0}
\end{figure}
%===============================================================================
The case   
$B=140$~MeV reaches $\sim 1.9 M_{\odot}$ but after the transition has set in,  
the star configurations get quickly unstable. These stars support, if they 
support, only a very small QM core and subsequently become unstable.\\
However, the transition pressure rises with the increase of $g_{\omega}$, which 
generates eventually an unstable QM core.
%====== = = = = = = = = = == = = = = = = = = = = = = = = = = = == = = = = = = = 
\begin{figure}[H]
\center
\includegraphics[width=\columnwidth]
{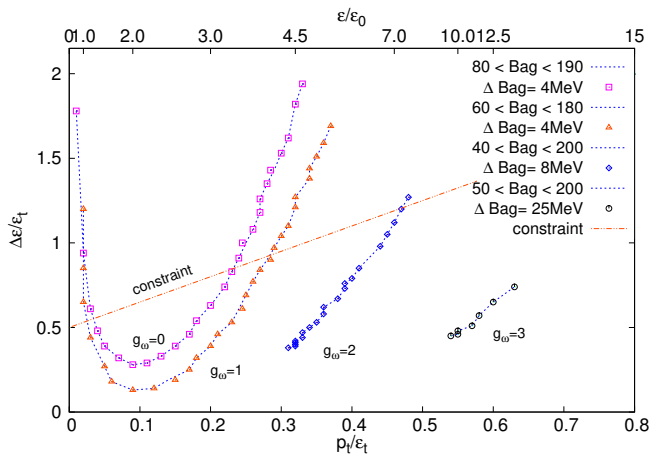}
\caption{\textit{Phase diagram for hybrid stars with fixed $g_{\omega}$ 
                 while varying $40 MeV \leq B \leq 200$~MeV at constant
                 $m_{\sigma}=600$~MeV and $m_q=300$~MeV. The axes display the 
                 transition pressure $p_t$ and the energy density discontinuity 
                 $\Delta \epsilon$ in units of the nuclear energy density at the
                 transition $\epsilon_t$.
    }}
\label{fig:constraintgw}
\end{figure}
%====== = = = = = = = = = == = = = = = = = = = = = = = = = = = == = = = = = = = 
The labelling of the axes in the phase diagram for fixed $g_{\omega}$ in 
fig.~\ref{fig:constraintgw} is the same as for fixed B in 
fig.~\ref{fig:constraintbag}. Generally, increasing the value of the repulsive 
coupling of the QM EoS leads to a higher $p_t$ and also a larger discontinuity 
$\Delta \epsilon$ for a given B. The higher the repulsive force within the QM 
core, the higher is $p_t$ for a QM core to appear. For the transition to occur 
at 2$\epsilon_0$, B has to be at least $104$~MeV in case of zero repulsion 
($g_{\omega}=0$), corresponding to the minimum of the plotted data in 
fig.~\ref{fig:constraintgw}. For $g_{\omega}=1$, B has to be at least $84$~MeV 
to be located at 2$\epsilon_0$. Both cases lead to stable hybrid star 
configurations, shown in figs.~\ref{fig:mr0} and \ref{fig:cp0} for 
$g_{\omega}=0$.\\ 
However, both trends are parabola like, crossing the constraint line twice, 
whereas the $g_{\omega}=2$ and the $g_{\omega}=3$ case stay below the constraint
(except for the choice $g_{\omega}=2$ and $B \gtrsim 190$~MeV). The 
$g_{\omega}=2$ case in the range $50 < B <200$~MeV  
corresponds to $4.5 \leq \epsilon_t \leq 7$. There a connected hybrid 
star branch, even if very small and hardly observable, exists up to 
$B \simeq 180$~MeV. The stars get unstable almost immediately after the 
appearance of the QM core. A higher value of B leads to transitions at already 
unstable mass-radius configurations. In case of even higher repulsion 
$g_{\omega}=3$ the transition takes place not before 10-14 times nuclear energy 
density at an already unstable mass-radius configuration.
% Hier unten noch was dazu schreiben!!!
Our results match the results from \cite{Alford:2014dva, Alford:2015dpa, Alford:2015gna}.\\
An investigation in the phase space by variation of $m_{\sigma}$ and $m_q$ lead
us to the conclusion that neither $\frac{\Delta \epsilon}{\epsilon_t}$ nor 
$\frac{p_t}{\epsilon_t}$ changes in an adequate amount to get a relatively
large jump in energy density accompanied with a small transition pressure, which
is an essential requirement for twin stars, see fig.~\ref{fig:twin_region}.
Their attractive character through varying both quantities is far weaker than 
the variation of $g_{\omega}$ and B 
\cite{Ko:1994en,Beisitzer:2014kea,Zacchi:2015lwa}.
  %=============================================================================
  \section{Comparison with other models}
  %=============================================================================
% Part von MH Anfang
% ====================================================
In the last section we have analyzed the parameter dependence of the resulting 
hybrid star properties within our HM-QM model. One main outcome of our analysis 
is the absence of a twin star region within the physical reasonable parameter space. 
Theoretically we have found a narrow parameter region where twin stars do exist 
($p_t/\epsilon_t < 0.05$), however, within all of these EoSs the HM to QM phase 
transition appears at irrelevant low density values ($\epsilon_t < \epsilon_0$). 
As the existence of twin stars have been found in many different kind of 
phase-transition scenarios, e.g. hadron-quark phase transition  
\cite{Mishustin:2002xe,Hanauske03,Bhatta04} (using a  
Maxwell- or Gibbs construction), hyperon phase transition \cite{SchaffnerBielich:2002ki}, pion \cite{Kampfer:1981yr}
and kaon condensation \cite{Banik04,Banik:2004fa}, the question arises, 
what the main reason is, that we do not find twins in our model? On the one 
hand, in all the existing twin star models, the relevant EoS parameter region 
where twins occur, is always narrow and a 'parameter fine-tuning' is needed to 
achieve an EoS which will result in a twin star behaviour. On the other hand, we 
have carefully analysed the allowed parameter space in the last section and did 
not find a twin star solutions where $\epsilon_t > \epsilon_0$.
%====== = = = = = = = = = == =  = = = = = = = = = = = = = = = ==================
\begin{figure}[H]
\center
\includegraphics[width=\columnwidth]
{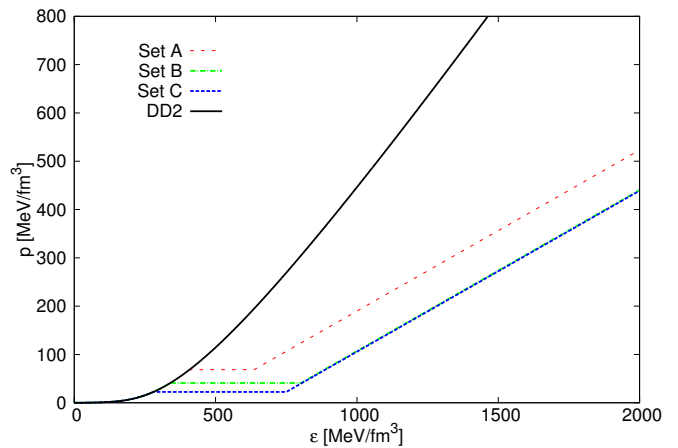}
\caption{\textit{The EoS for three different parameter sets corresponding to a
                 QM-EoS given by eq.~\ref{qmeostwo}. The parameters of the three
                 sets are displayed in table \ref{tab:twinpara}.}}
\label{fig:eos_mh}
\end{figure}
%===============================================================================
We show that the non-existence of twin stars in our model 
is due to the fact that the potential twin star area lies outside of our 
available parameter region and therefore cannot be reached in our simulations. 
By constructing the phase transition within our model we are 
not capable to choose arbitrary values for $\Delta \epsilon$, $\epsilon_t$ and 
$p_t$ (like Alford et. al \cite{Alford:2014dva, Alford:2015dpa, Alford:2015gna}), because we 
need to match the HM-EoS with the QM-EoS in a consistent way, i.e. find the
intersection between pressure p and chemical potential $\mu$ for the 
transition pressure $p_t$. 
%===============================================================================
\begin{table}[H]
\begin{center}
\begin{tabular}{|c|c|c|c|c|c|c|}
\hline\hline 
\textbf{Star sequence} & $p_t$/$\epsilon_t$ & $\Delta \epsilon$/$\epsilon_t$  & $M_1$ & $R_1$ & $M_2$ & $R_2$\\
\hline
$\bullet$ Set A & 0.168 & 0.56 & 1.69332 & 13.262 & 1.69794 & 11.722\\
\cline{2-7}
$\blacksquare$ Set B & 0.12 & 1.36 & 1.34586 & 13.208 & 1.26019 & 8.906 \\
\cline{2-7}
$\blacktriangle$ Set C & 0.08 & 1.68 & 0.96196 & 13.052 &  1.19709 & 7.893 \\ \hline\hline
\end{tabular}
\caption{\textit{The parameter choice for a constant speed of sound $c_s^2=\frac{1}{3}$
                 of the three different sets of star sequences with the respective
                 masses and radii of the corresponding branches (fig.~\ref{fig:mrr_mh})}}
\label{tab:twinpara}
\end{center}
\end{table}
% ====================================================
In this section we use the same density dependent DD2 EoS for the hadronic 
part, but we use a much simpler model for the QM sector. Similar to 
\cite{Alford:2014dva, Alford:2015dpa, Alford:2015gna} we take a QM-EoS, which is parameterized 
by the following three values: $p_t$, $\Delta \epsilon$ and
and $c_s$ (constant sound speed in quark matter) and which is 
given by eq.~\ref{qmeostwo}. In order to construct a 
comparable QM-EoS with respect to our model, we have used a fixed value of 
$c_s^2=1/3$ for the following calculations. The EoS for $p>p_t$ in this simple 
QM model has the following form \cite{Alford:2014dva,zdunik2013maximum}
\begin{equation}\label{qmeostwo}
p(\epsilon)=c_s^2 \left(\epsilon - \epsilon_* \right)\,\,,
\quad\hbox{with:}\,\, \epsilon_*:=\epsilon_t+\Delta\epsilon-\frac{1}{c_s^2}p_t\, ,
\end{equation}
where $\epsilon_*$ is the energy density at zero pressure. Fig.~\ref{fig:eos_mh} shows the resulting 
EoSs for three choices of the parameters, which are given in tab.~\ref{tab:twinpara}.
In contrast to our model the parameters can be chosen in such a way that twin stars appear in a 
physically meaningful region.
%====== = = = = = = = = = == =  = = = = = = = = = = = = = = = ==================
\begin{figure}[H]
\center
\includegraphics[width=\columnwidth]
{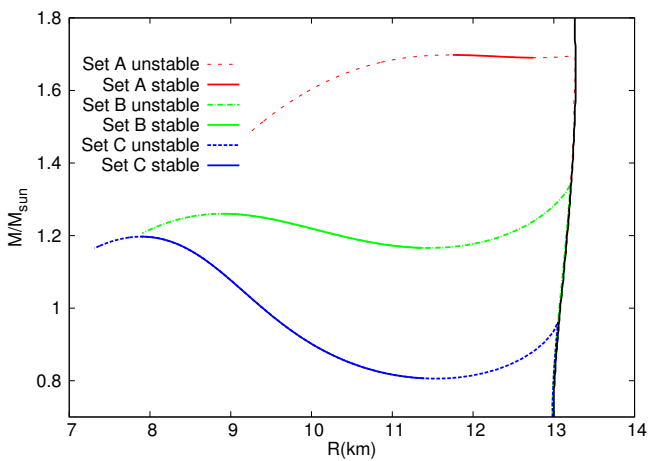}
\caption{\textit{The mass-radius relation for three different parameter sets 
                 corresponding to a QM-EoS given by eq.~\ref{qmeostwo}. The 
                 parameters of the three sets are displayed in 
                 tab.~\ref{tab:twinpara}. Set C shows impressive the appearance 
                 of a second stable branch, where $M_2 > M_1$: The maximum mass
                 of the second branch is larger than the maximum mass of the 
                 first branch. All displayed solutions are twin star solutions.
                 }}
\label{fig:mrr_mh}
\end{figure}
%===============================================================================
In Figs.~\ref{fig:mrr_mh} and \ref{fig:cp_mh} the mass-radius relations and the 
radius-mass curves of the three chosen representative twin star parametrizations 
are displayed. The 
Set A mass-radius relation has been calculated by using the parameter 
configuration: $\Delta \epsilon/\epsilon_t=0.56$ and $p_t/\epsilon_t=0.168$, 
which is located below the constraint-line given by 
eq.~\ref{eq:constraint} (see fig.~\ref{fig:twin_region}).
This configuration is located right at the corner of the twin star 
region boundary lines and the differences between the maximum masses of the 
first and second sequence is very small ($M^{max}_{1}=1.69332 M_{\odot}$ and 
$M^{max}_{2}=1.69794 M_{\odot}$).
Set B displays a twin star 
where the first sequence maximum mass lies above the maximum mass of the twin 
star ($\Delta \epsilon/\epsilon_t=1.36, p_t/\epsilon_t=0.12$).
The Parameter Set C curve shows the mass-radius relation of 
a twin star sequence with a rather high value of $\Delta \epsilon/\epsilon_t=1.68$ but a 
low value of $p_t/\epsilon_t=0.08$. The phase transition starts at low 
density and the maximum mass of the first sequence is much lower than the 
maximum mass of the twin star sequence (see table \ref{tab:twinpara}).
In this model too the neutron star sequence 
continuously moves to the hybrid star branch and hybrid stars with a tiny 
quark core are stable for a short period. The 
connected stable hybrid star branch is very small and difficult 
to recognize, as the hybrid stars get soon unstable after 
formation of the tiny quark core. Nonetheless twin stars somehow manage 
to restabilize again at a higher transition pressure.
%====== = = = = = = = = = == =  = = = = = = = = = = = = = = = ==================
\begin{figure}[H]
\center
\includegraphics[width=\columnwidth]
{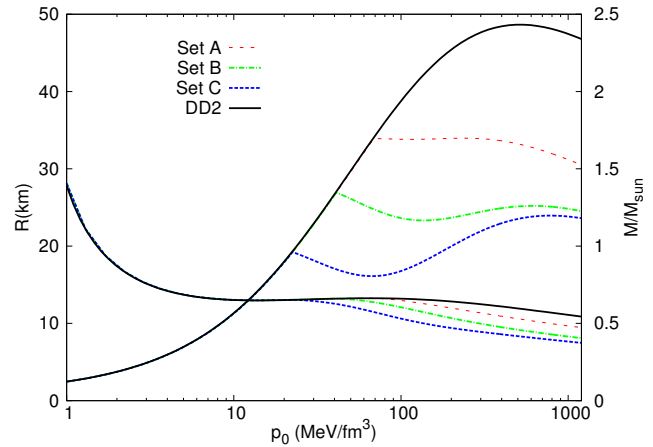}
\caption{\textit{The radius- and mass curves for three different parameter sets 
                 corresponding to a QM-EoS given by eq.~\ref{qmeostwo}. The 
                 parameters of the three sets are displayed in 
                 tab.~\ref{tab:twinpara}. Set C shows the appearance 
                 of a second stable branch, where $M_2 > M_1$. 
                 All displayed solutions are twin star solutions.
                 The curves starting in the upper left region
                 are the radius curves whereas the curves starting on the lower 
                 left side are the mass curves.}}
\label{fig:cp_mh}
\end{figure}
%===============================================================================
We do not get maximum mass values of the twin star configurations which are above 
the observational known value of $M=2.01M_\odot$, which means as a consequence, 
that all the twin star EoS are ruled out by nature.
%====== = = = = = = = = = == =  = = = = = = = = = = = = = = = ==================
\begin{figure}[H]
\center
\includegraphics[width=\columnwidth]
{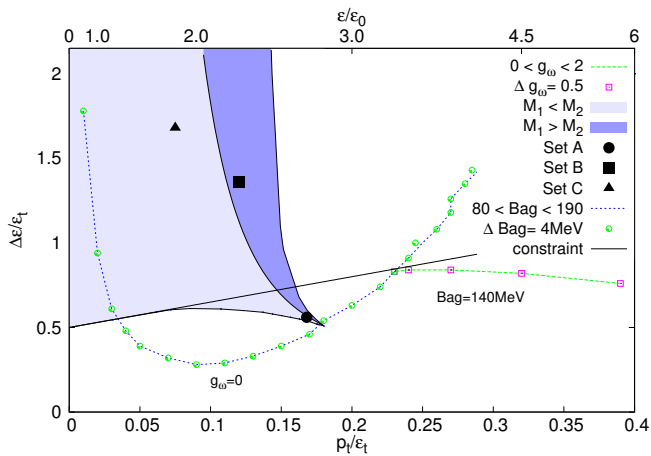}
\caption{\textit{Phase diagram for hybrid stars with the $g_{\omega}=0$-line and  
                 various values of B, and $B=140$~MeV at various $g_{\omega}$, both with
                 $m_{\sigma}=600$~MeV and $m_q=300$~MeV. The axes display the 
                 transition pressure $p_t$ and the energy density discontinuity 
                 $\Delta \epsilon$ in units of the nuclear energy density at the
                 transition $\epsilon_t$. The shaded regions displays the twin
                 star regions with either $M_1<M_2$ (light blue) or $M_1>M_2$ (blue)
                 calculated with the QM-EoS given by eq.~\ref{qmeostwo}.
                 The three twin star parameters sets are 
                 labelled as Set A, Set B and Set C.}}
\label{fig:twin_region}
\end{figure}
%===============================================================================
%===============================================================================
In Fig.~\ref{fig:twin_region} the twin star region in the model utilized by  
Alford et. al \cite{Alford:2014dva, Alford:2015dpa, Alford:2015gna} is compared with the 
space of available parameters within our model. 
It can be easily seen that the main part of the region where twin stars exist 
lies out of our attainable values of $\Delta\epsilon / \epsilon_t$ 
and $p_t / \epsilon_t$. Solely for irrelevant low values of $p_t / \epsilon_t$ we find 
a twin star area, see figure~\ref{fig:mr0}. The cusp at the lower end of the twin star region at 
($p_t / \epsilon_t = 0.18$, $\Delta\epsilon / \epsilon_t =0.51$) overlaps in a 
tiny region with the curve for $g_\omega=0$, however, we do not find any twin star in 
this parameter range. The radius-mass properties of hybrid stars near to the parameter 
region of the cusp, almost reach a twin-like structure (see e.g. the curve with $B=120$ 
MeV in fig.~\ref{fig:cp0}), though they never accomplish it mathematically. The reason 
for this apparent contradiction is causally determined in the fact that the sound speed 
for $g_\omega=0$ is not constant and slightly below the value which has been chosen to 
calculate the twin star area (see fig.~\ref{fig:sos}). As pointed out in 
\cite{Alford:2014dva}, a decrease of $c_s^2$ has the effect of scaling down the size of 
the twin star area and moves the cusp at the end of the twin star region 
upwards. Therefore, the absence of twin stars at the intersection of the cusp region is 
due to the energy dependence of the sound speed, which lowers its average value below 
$c_s^2=1/3$.
The line between the shaded areas separates whether 
the mass of the first branch $M_1$ lies above (blue) or below (lighter blue) 
the mass of the second stable branch $M_2$.
The $g_{\omega}=0$ line with $B=120-124$~MeV gets closest to the twin star
region.\\
Nevertheless, twin stars 
in general could exist in nature, as other models have been constructed 
\cite{Benic:2014jia,Dexheimer:2014pea} that satisfy the $M^{max}>2.01 M_\odot$ constraint. 
% Part von Matthias Hanauske- Ende
% ====================================================
  %=============================================================================
  \section{Conclusions}
  %=============================================================================
  % ==============================================================================
  % ==============================================================================
  % ==============================================================================
  % ==============================================================================
% From Macher and JSB pt_in_CS im paperi ordner
In this work we employ a density dependent hadronic matter EoS and a density dependent chiral quark matter EoS 
to find the phase transition from one phase to the other within compact stars.
The quarks couple to the meson fields via Yukawa-type interaction terms.
We utilize a Maxwell 
construction, i.e. assume that there is a sharp transition at a given transition pressure.
The transition pressure is identified when the chemical potential equals the pressure, from
that point on the dominance in the EoS flips and the larger energy density from the QM EoS prevails.   
Within our parameter range we found stable hybrid star solutions and investigated 
the relation of the QM core size and the appropriate stability of the star.
In the SU(3) Quark Meson model utilized for the QM EoS four parameters can be varied, 
from whom two of them ($m_{\sigma}$ and $m_q$) have little 
effect on the results. We conclude that a larger repulsive coupling $g_{\omega}$ 
and a larger vacuum pressure B do not allow for a large QM core to appear but
reach the $2 M_{\odot}$ limit, whereas small values of both quantities generate hybrid
star solutions with a corresponding, large QM core, but the configurations stay below the
$2 M_{\odot}$ constraint.
Hybrid stars with high transition pressures are hard to distinguish from pure hadronic stars because of the tiny QM core. 
An appearing QM core generates an additional gravitational pull on the 
hadronic mantle. If the cores pressure can counteract this extra pull, the star is 
stable. For a too large discontinuity in energy density the star gets unstable since the 
pressure of the core is not able to counteract the extra gravitational pull
\cite{Alford:2014dva, Alford:2015dpa, Alford:2015gna}.
% ==============================================================================
In \cite{Alford:2014dva} Alford et. al use hadronic EoSes based upon works from
Heberler et. al \cite{Hebeler:2010jx} and Shen et. al \cite{Shen:2011kr}.
Their QM EoS is density independent and is parametrized through $p_t$, $\epsilon_t$ and, assuming 
a constant speed of sound, $c_s^2$ . 
% The last feature shifts the jump in energydensity to volitional values!!??
They conclude that for stars with at least $2 M_{\odot}$ a larger $c_s^2$ is advantageous, whereas for
$c_s^2=1/3$ a larger region in the phase diagram for stars with $\geq 2 M_{\odot}$ is excluded, 
which as a consequence restricts the other parameters $p_t$ and $\epsilon_t$.  
% ==============================================================================
In a proximate work Alford et. al \cite{Alford:2015dpa} apply the constant speed of 
sound parametrization to a Field-Correlator-Method calculation. The corresponding EoS is  
equipped with an additive density 
independent $\bar{q}q$-potential, corresponding to our density dependent vector coupling constant, 
and with a vacuum energy density term including gluon 
condensate contributions, analogous to the Bag constant utilized in our approach. 
Vacuum energy density term and Bag constant are in both cases additive, i.e. density independent. 
In both works the allowed region in the phase diagram 
for hybrid stars with more than two solar masses is shifted to high transition pressures at several 
times nuclear energy density ($3.5 \leq \epsilon/ \epsilon_0 \leq 6.5$).
The family of the Field Correlator Method EoSes 
(varying the two above mentioned quantities) covers only a limited region 
in the phase diagram due to a nearly density independent speed of sound ($c_s^2 \simeq 1/3$), whereas in 
our approach we achieved high transition pressures assuming a higher vector coupling constant. 
This feature on the other hand raises the speed of sound up to values $c_s^2 \sim 0.6$, 
which would leave space in the phase diagram for the other parameters $p_t$ and $\epsilon_t$, only we had no direct influence on them.
% for $g_{\omega}=3$ (physically reasonable stars in our work lie within $0.3 \leq c_s^2 \leq 0.5$).
% These results are analogous to ours except that the speed of sound for physically reasonable stars
% in our work lies within $0.3 \leq c_s^2 \leq 0.5$.
% ==============================================================================
However, we confirm the results Alford et. al \cite{Alford:2014dva, Alford:2015dpa, Alford:2015gna}
obtained and investigate 
further why we were not able to find a third familiy (twin stars) of compact stars within a 
physically meaningful parameter region.
The conclusion is that the chances for twins are best when the transition pressure 
is relatively low and the energy density discontinuity on the other hand relatively high, 
then an appearing QM core does not destabilize the star immediately. 
Likewise it gets harder to achieve the $2 M_{\odot}$ regime.
But if the discontinuity in energy density is too large, 
the pressure of the QM core is unable to counteract the additional downward 
pull and the star configurations becomes unstable.
A future work could study the interplay between hadronic- and quark matter EoS in greater detail to 
work out how to achieve the appropriate proportions between pressure and discontinuity in energy density for twin stars.
% ==============================================================================
% ==============================================================================
% ==============================================================================
% ==============================================================================
% ==============================================================================
\begin{acknowledgements}
The authors thank Laura Tolos and David Blaschke for discussions during the initial stage 
of this project. Furthermore we want to thank Thorben Graf and Rainer Stiele 
for helpful suggestions during the whole project.
We gratefully acknowledge Sophia Han and Mark Alford for pointing out an inconsistency in the 
twin star area region of the previous version of our article and additionally want to 
thank them for further remarks.
AZ is supported by the Helmholtz Graduate School
for Heavy-Ion Research (HGS-HIRe) and the Helmholtz Research School for Quark Matter (H-QM).
MH gratefully acknowledges support from the Frankfurt Institute for
Advanced Studies (FIAS).
\end{acknowledgements}
%===============================================================================
\bibliography{all_new2}
\bibliographystyle{apsrev4-1}
\end{document}